# Pairing for Greenhorn: Survey and Future Perspective

Mahender Kumar, Satish Chand
School of Computer & System Sciences
Jawaharlal Nehru University, New Delhi, India
mahendjnu@gmail.com, Schand@mail.jnu.ac.in

**Abstract:** Pairing is the most powerful tool in cryptography that maps two points on the elliptic curve to the group over the finite field. Mostly cryptographers consider pairing as a "black box" and use it for implementing pairing-based cryptographic protocols. This paper aims to give the overview of pairing as simple as possible for greenhorn and those who are working and wish to work in the pairing. The paper gives the concrete background of pairing and recommends an appropriate pairing among different choices for constructing pairing-based cryptographic protocols. We also analyze the bandwidth and computational efficiency of pairing and submitting those pairing suitable for implementing a cryptographic protocol for lightweight devices. Additionally, we discuss the extension of bilinear pairing to tri-linear and multilinear pairing and discuss a few assumptions to check their feasibility to implement multilinear pairing.

## 1. Introduction

One of the notable discoveries in cryptography has been the invention of a public-key cryptosystem (PKC), developed by Diffie and Hellman [1], which is implemented in a group of prime order. It is a famous protocol to figure out the secret shared key $g^{AB}$ on given $g^A$ and $g^B$ shared by two unknown entities, respectively, in the system without interacting them previously. The security of the Diffie-Hellman protocol is defined by the Decision Diffie-Hellman (DDH) assumption. So DDH is beautiful, and one of its applications is the implementation of first public-key encryption, invented by the Rivest, Shamir, and Adleman [2]. RSA further contributed to implementing several protocols, like OpenPGP, S/MIME, and SecureShell, which have been developed based on the RSA cryptosystem. In 1985, ElGamal [3] presented the ElGamal encryption scheme defined over the cyclic group, and its construction is based on the Diffie-Hellman key exchange method. Diffie-Hellman constructed their protocol using a group, defined over a finite field $F_p$. The difficulty with $F_p$ is that the discrete log problem in $F_p$ is not as hard as we need for a secure system. A sub-exponential algorithm breaks the discrete log in $F_p$. So we need a relatively large prime in order to make a system secure against such an attack. We are currently using a prime that is about 2000-bits, and the recommendation is to use prime as much about 3000 bits, which are relatively large primes, causing the protocol to be slow.

The search for other groups has been going on for quite a while. Other groups have a hard discrete log problem in which we can try to run the Diffie-Hellman protocol. For example, extension fields, matrix groups, and class groups have been explored for running the Diffie-Hellman protocol. However, they have either an easy discrete log or a sub-exponential discrete log problem. It would also result in broad parameters and be somewhat inefficient, or they have a slow group operation, which would again result in a slow protocol. The first group that has turned out to be better than $F_p$, of course, has been the group of points on the elliptic curve over a finite field, proposed by Miller 1985 [4] and Koblitz 1985 [5]. The best-known algorithm for the discrete log takes exponential time, its square root of the relative prime size. Thus, we can use much smaller primes and achieve the same complexity as working in $F_p$. At present, we can use smaller primes because the discrete logarithm is much harder in this group with efficient group operations. As per the NIST guidance, a 3092-bit modulus in $F_p$ achieves a 128-bit symmetric key security, while for the same level of security, we need a 256-bit key in an elliptic curve, shown in Table 1. Hence, it is relevant for those applications where smaller bandwidth and memory are required.

Table 1. NIST recommended key size

| Security strength | Key size (bits) | |
|---|---|---|
| k | ECC | RSA |
| 80 | 160 | 1024 |
| 112 | 224 | 2048 |
| 128 | 256 | 3072 |
| 192 | 384 | 7680 |
| 256 | 512 | 15360 |

For many years, the impressive quality of the elliptic curve is that people have only looked at it for their efficiency improvements. It turns out that elliptic curves actually have a supplementary fabrication called a pairing that gave birth to a new area of cryptography, called pairing-based cryptography (PBC). A pairing abstractly is something that operates on two source groups G and resultant on another target group called $G_T$. Generally, the source group consists of the points on an elliptic curve, and the target group ($G_T$) is just an element in a finite field. It takes two points in the source group and maps them to the target group in such a way that the exponents multiply.

The pairing has recently gained much attention, which is being standardized as a next-generation cryptosystem. It maps the discrete logarithm in a subgroup of an elliptic curve to the discrete logarithm in a finite field. Menezes et al. [6] present the Weil pairing and attack on pairing to efficiently reduce the elliptic curve-based discrete logarithm problem (ECDLP) to a discrete logarithm problem (DLP) over a finite field using Weil pairing. This attack is known as the MOV attack, based on the initials of three inventors. Similarly, Frey and Ruck [7] use the

Tate pairing to reduce the ECDLP to DLP in a finite field, known as FR attack. It is well-known that not all elliptic curves are suitable for pairing. Thus, it is challenging to find the suitable elliptic curves containing a subgroup with optimal embedding degree k. Here, k is big enough to secure against FR attack but small enough such that the arithmetic in a finite field is efficiently computable. There has been an in-depth study of the elliptic curves suitable for pairing, known as pairing-friendly elliptic curves. Generally, they have a sizeable prime order subgroup with a small embedding degree.

*Contribution.* This article is intended to provide a sketch of pairing for the novice and non-cryptographer who wish to use pairing for a cryptographic protocol. Currently, most of the cryptographers utilize pairing as a "black-box" and construct protocols by considering some pairing properties. However, it is good to assume pairing, as black-box as it allows completely ignore the mathematical behind pairing and mainly focus on the utterly cryptographic angle of research as the details of pairing, such as their selection and implementation is significantly complicated. This article gives an overview of different pairing and distinct methods that pairing can be implemented, along with the features of each pairing, in terms of their applicability. It attempts neither to give new research nor to propose the problems for further study. Our goal is to present in a simple way to understand the pairing, the criteria that need to keep in kind while selecting the different choice of pairing, and designing system based on pairing. Another objective is to discuss the bandwidth and computation efficiency recommendations for lightweight devices. Besides, we also discuss the future perspective of pairing and found that multilinear map can be a powerful tool for implementing protocol. However, the security of the cryptographic protocols based on such multilinear maps is not clear.

*Organization.* The remaining article is organized as follows. We provide an introductory background on the elliptic curve and pairing that could be understandable to a learner in section 2. Section 3 gives distinct pairings and recommendations for implementing cryptographic protocols. We then give the more refined analysis for choosing effective pairing for lightweight devices, in section 4. Section 5 discusses the feasibility of a multilinear map as a future perspective of pairing. Finally, we give a conclusion in section 6.

## 2. Background

### 2.1. Elliptic curves

The elliptic curves have a vibrant and attractive history in the last century, which has been studied by several mathematicians. They have solved numerous wide ranges of mathematical problems, for example, Fermat's last theorem. One of the advantages of elliptic curve is its ability to construct a group structure that makes it possible to implement the cryptographic protocols. The difficulty of finding the discrete logs in a group, i.e., the absence of a sub-exponential time algorithm, is the space that the elliptic curves have over the system based on the multiplicative group for the finite field. In fact, the ECC is efficient and more secure tools that has been used for constructing cryptographic protocols for systems having limited computation and storage with same level of security.

Consider a field K and its algebraic closure $\overline{K}$. The Weierstrass equation ($y^2 + a_1xy + a_3y = x^3 + a_2x^2 + a_4x + a_6$) over field K, where $a_i \in$ K and $1 \leq i \leq 6$ is basis of elliptic curve. Thus, an elliptic curve E: $y^2 = x^3 + ax + b \subset K^2$ is a non-singular[1] projective closure of the smooth affine curve, where $a, b \in$ K. Indeed, elliptic curve is derived from the Weierstrass equation, whose characteristic[2] of field K is not equal to 2 or 3. There must also be a projective point O:= (0 : 1 : 0), as the point at infinity. Due to this, researchers consider the elliptic curves as affine curves and view the point at infinity individually that point plays an essential role in the implementation of modern cryptography. To find the non-singularity of the curve defined by the Weierstrass equation, we need to compute the value of the discriminant $\Delta = -16(4a^3 + 27b^2)$. An elliptic curve E is said to be non-singular if $\Delta \neq 0$. Let $a_i \in K \subset \overline{K}$, $\overline{K}$ is algebraic closure of K, which states that the curve is defined over K and can be written as $E/K$, where we define $E(K)$ as the set of K-rational points of E.

*2.2. Group structure of Elliptic curves*

To construct a secure cryptographic protocol, the points on an elliptic curve must form an abelian group structure. Let E/K denotes an elliptic curve E over K and P = $(x_P, y_P)$ and Q = $(x_Q, y_Q)$ are two points on E/K. The points on elliptic curve E/K constitute an abelian group, including the identity element denoted as O, the negation of point P is defined by $-P = (x_P, -y_P)$ and addition operation on P and Q such that $P \neq -Q$ is defined by $P + Q = (x_{P+Q}, y_{P+Q})$, where

$$x_{P+Q} = (\mu^2 - x_P - x_Q) \tag{1}$$

$$y_{P+Q} = (\mu(x_{P+Q} - x_P) - y_P) \tag{2}$$

and, μ is the slope defined in Equation (3),

$$\mu = \begin{cases} \left(\frac{y_Q - y_P}{x_Q - x_P}\right) & if\ P \neq Q \\ \left(\frac{3x_P^2 + a}{2y_P}\right) & if\ P = Q \end{cases} \tag{3}$$

It is to note that the addition operation in E/K is commutative and E/K is closed under addition and the additive inverse. According to B´ezout's Theorem, the line passing through the points $P$ and $O$ meets on $E/K$ at a unique point, which we denote $-P$. The *additive inverse* of point $P$ is a mirror of point $P$ about the x-axis. The line through $P$ and $Q$ intersects at the third point, denoted as $-(P + Q)$ on curve $E/K$ and the additive inverse of point $-(P + Q)$, i.e., $(P + Q)$ is known as *addition operation* on points on $P$ and $Q$ on elliptic curve. Thus, the addition

---

[1] Non-singular means that the graph has no cusps, self-intersections, or isolated points.
[2] The smallest number of 1s that sum to 0 is called the **characteristic** of the **finite field**, and the **characteristic** must be a prime number

operation is defined as: given three non-zero aligned points on the elliptic curve, $P + Q - (P + Q) = O$. Consider a point $P \in E/K$, the line passing through $P$ is the tangent, and the addition of point P to itself is known as the *point doubling*. For any two points P and Q on elliptic curve E/K, there exists an element $a \in K$, such that $Q = aP$. The operation of adding the point $P$ to itself using the group addition law is known as the scalar point multiplication. The smallest positive integer $q$ such that $qP = O$ is known as the order of point $P$.

### 2.3. Elliptic curves over finite field

A finite field is a set of finite number of elements, for example, the set of integer modulo prime number $p$, denoted as $Z_p$ or $F_p$ or $GF(p)$ is a finite field. The finite field $F_p$ has $p$ elements from 0 to $(p-1)$. The order must be prime, otherwise it does not have the multiplicative inverse of each integers which is a necessary condition to form a field. The elliptic curve E over finite field $F_p$, $E(F_p)$ holds the required properties to form an abelian group structure. For larger prime, schoof algorithm compute the order of the prime field in $Olog(p)$ time. Let $P$ and $Q$ be the points on elliptic curve $E(F_p)$ such that $Q = aP$, $a \in F_p$. In this scenario, $P$ is generator of group, $Q$ is considered as public key and $a$ is known as private key. The ECC security is defined by the discrete logarithm problem on elliptic curves (ECDLP), that is to compute $a$ on given P and Q. The ECDLP creates a platform for constructing the asymmetric key algorithms, for example, elliptic curve digital signature algorithm (ECDSA), elliptic curve Diffie-Hellman key exchange (ECDHE) protocol and elliptic curve ElGamal based encryption.

### 2.4. Divisor on elliptic curve

A torsion group in an elliptic curve plays a primary role in understanding the concept of pairing-based cryptography. For an elliptic curve E, we denote the q-torsion group by $E[q]$, that is defined as $E[q] = \{P \in E(K)[q] | [q]P = 0\}$. Here, the point $P$ is said to be a point of finite order or torsion point. It can be seen that if $P, Q \in E[q]$ then $P + Q \in E[q]$ and $-P \in E[q]$. Thus, $E[q]$ is said to be a subgroup of $E$. Here, $P = E(K)[q]$ is represented as the point $P$ lying on the particular field $K$, like $Q$ or $R$ and $F_p$.

Suppose $f(x, y)$ is a rational function of two variables $x$ and $y$. There exist some points on E where the numerator of $f$ vanishes, and some points of $E$ where the denominator of $f$ vanishes. That means, $f$ has zeros and poles on $E$. One way to keep track of the zeroes and poles of a function is to compute the divide of f, where poles are related to projective coordinates. The divisor associated with $f$ is defined as the formal sum, given in Equation (4).

$$D = div(f) = \sum_{P \in E} n_P[P] \qquad (4)$$

where, coefficient $n_P \in Z_q$, in which many of them are nonzero, so $D$ is the finite sum. The degree of divisor is defined as the sum of its coefficients, as given in Equation (5).

$$\deg(D) = \deg\left(\sum_{P \in E} n_P[P]\right) = \sum_{P \in E} n_P \qquad (5)$$

Now, we define the sum of divisor, as given in Equation (6).

$$\text{sum}(D) = \text{sum}\left(\sum_{P \in E} n_P[P]\right) = \sum_{P \in E} n_P P \tag{6}$$

Here, $n_P P$ is the additive operation on P to itself $n_P$ times. One of the reason to use the divisor in pairing is, if we have a devisor we can reconstruct the function. Thus, the devisors are main ingredient to specify the pairing function in pairing-based cryptography, which is computed by the Miller algorithm.

*Miller Algorithm.* The Miller's algorithm maps two points on an elliptic curve to an element of a finite field. Suppose we have two points on the elliptic curve, says $P$ and $Q$, then the Miller algorithm computes elemet $x$ such that, $x \leftarrow e(P, Q)$ on the finite field. Recall that the points addition (e.g., $R \leftarrow P + Q$) and scalar multiplication operations (e.g. $R \leftarrow kP$) on the elliptic curve are equivalent to the elements multiplication (e.g., $R \leftarrow kP$) and exponentiation operations (e.g. $R \leftarrow p^k$) on the field, respectively.

Let $a \geq 1$ be an integer, which can be written in binary expansion as $a = a_0 2^0 + a_1 2^1 + \cdots + a_{n-1} 2^{n-1}$, where, $a_i \in \{0,1\}$ and $a_{n-1} \neq 0$. The Miller algorithm gives the function $f_p$ whose divisor satisfies $div(f_p) = q[P] - [qP] - (q-1)[O]$. One of the useful properties of the Miller's algorithm is bilinearity. Consider four points $P, Q, A$ and $B$, and two integers $a$ and $b$, where $A \leftarrow aP$ and $B \leftarrow bQ$. The Miller algorithm computes $x \leftarrow e(P, Q)$ for $P$ and $Q$, and $y \leftarrow e(A, B) = e(aP, bQ)$ for $A$ and $B$, which are related by $y = x^{ab}$. In other words, the bilinearity property is defined by Equation (7).

$$e(aP, bQ) = e(P, Q)^{ab} = e(bP, aQ) \tag{7}$$

Due to this bilinearity property, many cryptographic protocols have been constructed, which could never be possible or too complicated, such as short signature, identity-based encryption, and attribute-based encryption.

| Algorithm 1: Miller Algorithm |
|---|
| **Set** $T = P$ and $f = 1$ |
| **Loop** $i = n - 2$ down to 0 |
|     **Set** $f = f^2 \cdot g_{T,T}$ |
|     **Set** $T = 2T$ |
|     **If** $a_i = 1$ |
|         **Set** $f = f \cdot g_{T,p}$ |
|         **Set** $T = T + P$ |
|     **End If** |
| **End** i loop |
| **Return** value $f$ |

Where

$$g_{P,Q} = \begin{cases} \frac{y-y_P-\mu(x-x_P)}{y+x_P+x_Q-\mu^2} & if\ \mu \neq \infty \\ x - x_P & if\ \mu = \infty \end{cases} \quad (8)$$

and μ is slope.

## 2.5. Weil pairing

Let $P, Q \in E[q]$, and $f_P$ and $f_Q$ denote the rational functions on E[q] satisfying $div(f_P) = q[P] - q[0]$ and $div(f_Q) = q[Q] - q[0]$. The Weil pairing $e_W$ of $P$ and $Q$ is given by

$$e_W(P,Q) = \frac{f_P(Q+S)}{f_P(S)} \bigg/ \frac{f_Q(P-S)}{f_Q(-S)} \quad (9)$$

where $S \in E[q]$ is a random point such that all elements on right hand side are defined and nonzero, i.e., $S \notin \{O, P, -Q, P-Q\}$. The Weil pairing satisfies the following properties

- $q^{th}$ *root of unity*: For all $P, Q \in E[q]$, the weil pairing is $e_W(P,Q)^q = 1$.
- *Bilinearity*: For $P, P_1, P_2, Q, Q_1, Q_2 \in E[q]$, $e_W(P_1 + P_2, Q) = e_W(P_1, Q)e_W(P_2, Q)$ and $e_W(P, Q_1 + Q_2) = e_W(P, Q_1)e_W(P, Q_2)$.
- *Alternating*: For all $P, Q \in E[q]$, the weil pairing is $e_W(P, Q) = 1$, which implies that $e_W(P, Q) = e_W(Q, P)^{-1}$
- *Non-degeneracy*: For all $P, Q \in E[q]$, if $e_W(P, Q) = 1$, then $P = 0$.

## 2.6. Tate pairing

Tate pairing is computationally more efficient than the Weil pairing, not only because if need one call of Miller's algorithm, instead of two and have some optimization. Suppose there is an elliptic curve E over $F_q$, of order prime $l$, and let two points point $P, Q \in E(F_q)[l]$. Pick a rational function $f_P$ on E with divisor $div(f_P) = l[P] - l[O]$. The Tate pairing of $P$ and $Q$ is computed as

$$\tau(P,Q) = \frac{f_P(Q+S)}{f_P(S)} \in F_q \quad (10)$$

where $S \in E(F_q)$ is point such $f_P(Q+S)$ and $f_P(S)$ are defined and nonzero.

Let an elliptic curve over $F_p$ be $E(F_p)$ and $r \geq 1$ be an integer such that $p$ is not divisible to $r$. Then, the embedding degree is defined as the smallest integer $k$ such that

$$E(F_{p^k}) \cong Z_r \times Z_r \quad (11)$$

The embedding degree helps the Weil pairing to map DLP on the elliptic curve $E(F_p)$ into the DLP in the field $F_{p^k}$.

The MOV algorithm [6] reduces the ECDLP in elliptic curve $E(F_p)$ to the DLP in field $F_{p^k}$. Algorithm 2 summarizes the MOV algorithm.

| **Algorithm 2: MOV algorithm** |
|---|
| 1. Compute $N = \#E(F_{p^k})$ |

> 2. Pick any point $T \in E(F_{p^k})$ but $T \notin E(F_p)$.
> 3. Let $T' = (N/l)T$. If $T' = O$, run step 2. Otherwise $T'$ is point of order $l$, so run step 4.
> 4. Compute Weil pairing as
>    $a = e_W(P, T') \in F_q$ and $b = e_W(Q, T') \in F_q$
> 5. Solve the DLP for $a$ and $b$ in $F_q$, find an exponent $n$ such that $b = a^n$.
> 6. Also $Q = nP$, so the ECDLP has been solved.

*2.7. Modified Weil pairing*

Let an elliptic curve over $F_p$ be $E$ and $l \geq 3$ be a prime. Let a point $P \in E(F_q)[l]$ on elliptic curve of order $l$ and $\varphi : E \rightarrow E$ be map from E to itself. Then the mapping $\varphi$ is said to be $l$-distortion map for $P$ if it satisfies two properties:

- $\varphi(nP) = n\varphi(P)$ for all $n \geq 1$.
- $e_W(P, \varphi(P))$ has $l^{th}$ root of unity. That means, for any integer $r$ multiple of $l$, $e_W(P, \varphi(P))^r = 1$.

Let an elliptic curve over $F_p$ be $E$ and $l \geq 3$ be a prime. Let point $P \in E(F_q)[l]$ on elliptic curve order $l$ and $\varphi$ be an $l$-distortion map for $P$. The modified Weil pairing $e_W$ on $E[l]$ is defined by $e_W(Q, Q) = e_W(Q, \varphi(Q))$. Non-degeneracy is the important property of the modified Weil pairing.

*2.8. Mathematical assumptions*

Let discuss about some mathematical assumptions. Basically, if we have a pairing then the Decision Diffie-Hellman problem turns out to be easy. For a given $P, Q, R, S \in E(F_p)$ and $\forall\, x, y, z \in Z_q$, such that $Q = xP$, $R = yP$ and $S = zP$, DDH is to test whether $z = xy$ [8]. But using pairing, it easy to break DDH problem: take two pairings $e(P, zP)$ and $e(xP, yP)$ and by the property of pairing we get the a equality only if z is equal to x time y, i.e., $e(P,P)^z = e(P,P)^{xy}$. Thus, DDH becomes a breakable problem for pairing and we can use it for pairing groups. The other thing is that we obtain this reduction from discrete log in G to discrete log in target group $G_T$, otherwise this discrete log in source group will not be hard.

So, there are the two immediate properties: bilinearity and degeneracy that we get out of pairing. Now, we discuss the complexity assumptions that come up with it. We get the standard complexity assumptions that we know and the discrete log is a perfectly kind of a requirement for pairing-based groups. We need the Computational Diffie-Hellman (CDH) problem. For a given $P, Q, R \in E(F_p)$ and $\forall\, x, y \in Z_q$, such that $Q = xP$, and $R = yP$, CDH problem is to compute xyP. Actually, CDH is not hard, it's an easy problem. We just add one more element to CDH to replace it with a new assumption, Bilinear Diffie-Hellman (BDH) problem [9]. For a given $P, Q, R, S \in E(F_p)$ and $\forall\ x, y, z \in Z_q$, such that $Q = xP$, $R = yP$ and $S = zP$, BDH is to finding $e(P,P)^{xyz} \in F_q$. Several encryption schemes actually use this as randomizer for encryption.

## 3. Consequence of pairing

In general, the basic form of pairing is $e: G_1 \times G_2 \rightarrow G_T$. Here, $G_1$, $G_2$ and $G_T$ are the cyclic group of prime order q. for the purpose of this paper, we defined three basic types of pairing,

- Type 1: $G_1 = G_2$; symmetric pairing.
- Type 2: $G_1 \neq G_2$; asymmetric pairing, and there must be an efficiently computable homomorphism $\Phi: G_2 \rightarrow G_1$.
- Type 3: $G_1 \neq G_2$; asymmetric pairing, but there is not found any efficiently computable homomorphism $\Phi: G_2 \rightarrow G_1$.

This section explore those issues due to which the selection of groups and pairings. We consider that the group $G_1$, $G_2$ and $G_T$ and pairing e(.,.) constitutes a system parameter of any pairing-based cryptographic protocol. In each case, there exists a homomorphism between $G_1$ and $G_2$, but computing such homomorphism is as difficult as computing discrete logarithms in the groups. If there an efficient homomorphism from $G_2$ to $G_1$ but not from $G_1$ to $G_2$, we call such pairing as Type 2 pairing. The environment where homomorphism in both direction are efficiently computable, we call such pairing as Type 1 pairing. If there is no such efficient computable homomorphism exists in any direction, we call it Type 3 pairing.

This pairing type distinction is relevant for the design of cryptographic protocols. There have been discussed many pairing-based cryptographic protocols whose security proof does not apply if the protocol is constructing on the pairing of Type 3. Galbraith et al. [10] discuss four assumptions about pairing when they are using for implementing the cryptographic protocols.

- Hash to $G_2$
- Short representation for $G_1$ elements
- Efficient computable homomorphism from $G_2$ to $G_1$
- Generating system parameters that achieve at least k bit of security in time polynomial k.

Several researchers believe that few properties can be easily satisfied, but it is not easy to simultaneously achieve all these properties. The Type 1 pairing is implementing using supersingular curves over fields of characteristic 2 or 3 and of large prime characteristic. The Type 2 pairing is implementing on ordinary curves, and the homomorphism mapping from $G_2$ to $G_1$ is traceable. The Type 3 pairing is on ordinary cures and $G_2$ is typically taken to be the kernel of trace map. Galbraith presented that there have some implementations of pairing for which the property holds for all security levels. It has to be noted that it is difficult to achieve all features simultaneously.

## 4. Recommended pairing for lightweight devices

Nowadays, emerging technologies such as IoT and sensor networks use constrained devices, such as sensor node, RFID tag, and smartcard, that lack the CPU and memory capacity to run traditional cryptographic protocols. To achieve security in such an environment, we construct a secure cryptographic protocol with low-overhead requirements.

Recently, PBC is becoming popular in achieving tight security in a resource-constrained environment. One of PBC's practical applications, i.e., an identity-based cryptosystem, is considered an appropriate method in this type of platform. PBC is highly valuable to construct non-interactive key agreement schemes that allow two nodes to share a shared session key without interacting with others. Since sensor networks are highly mobile, the identity-based non-interactive key exchange is applicable in WSN.

As we know, pairing operation, on comparing with ECC-based operations and traditional cryptographic operations, is computationally inefficient. Many cryptographers assume that pairing can easily be implemented in a resources-constrained environment. However, it is not true that it is directly run on sensors. Pairing is used for resource-constrained devices with some assumptions and optimizations. We now discuss such assumptions, consideration, and optimization to use pairing for devices with the least computation and storage capability.

We now focus on the following optimization during the construction of pairing-based cryptographic protocols when they are designing for constrained-devices.

- Eliminate unnecessary costly infrastructure.
- Eliminate interaction between two nodes during key sharing.
- Designing a protocol in such a way that costly operations, such as pairing is run on the base station.
- Use a signature scheme that outputs the least signature size.
- Preprocessing the costly independent operations.
- Use efficient pairing, such as TinyPairing, TinyPBC and TinyECC, for sensor node, if necessary.

Identity-based encryption (an application of PBC) avoids PKI, which is required in traditional PKC. Instead, it uses the identification of the node as a public key for encryption. Besides, the integration of PBC and IBC supports a non-interactive key exchange protocol. Due to limited storage and bandwidth, they require a short signature. BLS short signature, one of application of PBC provide a signature of order of 160-bit level of security identical to that of 320-bit DSA signature. Several authors are adopting an online/offline technique that helps in preprocessing the costly independent operations once and stored in node memory before installation. Currently, many authors use TinyTate, TinyPBC and TinyPairing for constructing a protocol for lightweight devices.

Tate pairings is more efficient for fast computation than Weil pairing. Oliveira et al. [11] show that Tate pairing-based protocols are feasible in resource-limited nodes and presented *TinyTate* that takes 30.21seconds computation time with 1831bytes of RAM and 18384bytes of ROM.

Later, Barreto et al. [12] leads to an Eta-t ($\eta_T$) pairing which is superior among Eta generalization. Table 2 shows the implementation of Tate and eta pairing in four wireless sensor nodes: MICA2, TmoteSky, Imote2 (13MHz), and Imote2 (104MHz) run on 8-bit ATmega128L. It has been clear that eta ($\eta_T$) pairing shows the least computation overhead and consumes less memory space compared with Tate pairing.

Xiaokang et al. [13] proposed *Tiny pairing* in the field of embedding degree 3 which uses new cubing algorithm with faster modular reduction and a faster polynomial multiplication. Tiny pairing computation is faster than Eta-t pairing and other implementations, which 154 bytes RAM and 8576 bytes ROM. Table 3 compares the implementation of different pairings on MICAz.

Table 2. Pairing implementation on four different sensor nodes [14].

| Nodes | Pairing | Timing (sec) | ROM (KB) | Stack (KB) | Current draw (mA) | Energy usage (mJ) |
|---|---|---|---|---|---|---|
| Mica2 | Tate | 7.43 | 60.91 | 3.39 | 7.86 | 62.73 |
| | eta-t | 2.66 | 47.91 | 3.17 | 7.88 | 175.65 |
| Tmote sky | Tate | 4.61 | 34.9 | 3.39 | 3.45 | 17.7 |
| | eta-t | 1.71 | 23.66 | 4.17 | 3.68 | 50.9 |
| Imote2 (13Mhz) | Tate | 0.62 | 44.4 | 3.75 | 31 | 12.12 |
| | eta-t | 0.46 | 29.5 | 4.12 | 31 | 16.34 |
| Imote2 (104Mhz) | Tate | 0.08 | 44.4 | 4.12 | 66 | 3.76 |
| | eta-t | 0.06 | 29.55 | 3.75 | 66 | 5.02 |

Table 3. Implementation of different pairing on MICAz nodes.

| Property | Tiny Tate [11] | Tiny PBC [15] | Tiny Pairing [13] |
|---|---|---|---|
| **Pairing type** | Tate | Eta-t | Eta-t |
| **Embedding degree** | 2 | 4 | 3 |
| **Security level** | 40 | 80 | 80 |
| **Pairing Time (sec)** | 30.21 | 5.45 | 5.3 |
| **ROM** | 18.38 | 47.5 | 8.57 |
| **RAM** | 1.83 | 0.37 | 0.154 |
| **STACK** | NA | 2.87 | NA |

## 5. Multi-linear pairing

Pairing (bilinear) map has indeed found a great mathematical tool in cryptography. Cryptographers have pointed out the extension of the degree of linearity, such as trilinear map, k-linear map, would provide, even, a more robust application in cryptography. An open problem in pairing is to extend the bilinear pairing to construct a secure tri-linear and multilinear map. Thus, constructing multilinear map is to construct the group G and $G_T$ in such a way that the discrete log in G is difficult. There must be an efficiently possible computable non-degenerate n-linear map $e: G^n \to G_T$. This construction can give robust solutions to the existing problems and a new functional and homomorphic encryptions schemes.

Boneh and Silverberg [16], in 2003, gave two interesting applications of multilinear maps, such as multi-partite Diffie-Hellman key exchange and efficient broadcast encryption. One could say that multilinear maps are at the heart of indistinguishability obfuscation, witness encryption, multiparty key exchanges, among many other applications. However, it was not easy to construct such maps from the realm of algebraic geometry, until 2013, when Garg et al. [17] gives the first reasonable construction of multilinear maps on lattices (GGH13). Later, Coron et al. [18] gave a different construction (CLT13) that relied on the hardness of integer factorization and gave a practical one-round 7-party Diffie-Hellman key exchange protocol. Later Gentry et al. [19] discuss a different graph-induced multilinear encoding scheme from lattices (GGH15).

Unfortunately, these constructions do not base on successful mathematical assumptions. There have found several attacks, which illustrate that they are incapable of achieving desirable security requirements and could not use to construct secure protocol. Till now, the security of the cryptographic protocols based on such multilinear maps is not clear.

## 6. Conclusion

This paper presented an overview of pairing as simple as possible for beginners. It also supports to opt an appropriate pairing among different choices for constructing pairing-based cryptographic protocols. The paper is intended for the beginners to cryptography to understand the elliptic curves and pairing on them. We also discuss the assumptions that to be considered for constructing the protocols. Additionally, we describe the bandwidth and computational efficiency of pairing and recommend some optimization while constructing pairing-based protocols for lightweight devices. Further, we discuss the extension of bilinear pairing to tri-linear and multilinear pairing and conclude that the security of the cryptographic protocols based on such multilinear maps is not clear.

## 7. Declaration

*Availability of data and materials*. This research did not used any dataset.

*Funding*. The authors declares that they have no known financial interest or personal relationships that could have appeared to influence the work reported in the paper.

*Acknowledgements*. The authors would like to thank the anonymous reviewers.